\documentclass[prl,twocolumn,notitlepage,superscriptaddress,showkeys,showpacs]{revtex4-1}
\usepackage{amsmath,amsfonts,amssymb,amsthm,epsfig,array}
\usepackage{dsfont}
\usepackage{graphics}
\usepackage{float}
\usepackage{bm}% bold math
\usepackage{verbatim}%\begin{comment} ... \end{comment}
\usepackage{gensymb}
\usepackage[dvipsnames]{xcolor}
\usepackage[hidelinks,colorlinks,linkcolor=blue,
citecolor=blue,urlcolor=blue]{hyperref}
\usepackage[titletoc,title]{appendix}

\newcommand{\sign}{{\rm sign}}

\newcommand{\beq}{\begin{equation}}
\newcommand{\eeq}{\end{equation}}
\newcommand{\beql}{\begin{equation*}}
\newcommand{\eeql}{\end{equation*}}
\newcommand{\beqn}{\begin{eqnarray}}
\newcommand{\eeqn}{\end{eqnarray}}

\begin{document}

\title{Lattice distortion induced first and second order topological phase transition in rectangular high-T$_{\rm  c}$ superconducting monolayer}
\author{Li Chen}
\thanks{These authors contribute equally to this work.}
\affiliation{School of Physics, Huazhong University of
Science and Technology, Wuhan, Hubei 430074, China}
\affiliation{Wuhan National High Magnetic Field Center, Huazhong University of
Science and Technology, Wuhan, Hubei 430074, China}
\author{Bin Liu}
\thanks{These authors contribute equally to this work.}
\affiliation{School of Physics, Huazhong University of
Science and Technology, Wuhan, Hubei 430074, China}
\affiliation{Wuhan National High Magnetic Field Center, Huazhong University of
Science and Technology, Wuhan, Hubei 430074, China}
\author{Gang Xu}
\email{gangxu@hust.edu.cn}
\affiliation{Wuhan National High Magnetic Field Center, Huazhong University of
Science and Technology, Wuhan, Hubei 430074, China}
\affiliation{School of Physics, Huazhong University of
Science and Technology, Wuhan, Hubei 430074, China}
\author{Xin Liu}
\email{phyliuxin@hust.edu.cn}
\affiliation{School of Physics, Huazhong University of
Science and Technology, Wuhan, Hubei 430074, China}
\affiliation{Wuhan National High Magnetic Field Center, Huazhong University of
Science and Technology, Wuhan, Hubei 430074, China}
\date{\today}

\begin{abstract}

We theoretically study the lattice distortion induced first and second order topological phase transition in rectangular FeSe$_{x}$Te$_{1-x}$ monolayer. When compressing the lattice constant in one direction, our first principles calculation shows that the FeSe$_x$Te$_{1-x}$ undergoes a band inversion at $\Gamma$ point in a wide dopping range, say $x\in(0.0,0.7)$, which ensures coexistence of the topological band state and the high-T$_{\rm c}$ superconductivity. This unidirectional pressure also leads to the C$_4$ symmetry breaking which is necessary for the monolayer FeSe$_x$Te$_{1-x}$ to support Majorana corner states in the either presence or absence of time-reversal symmetry. Particularly, we use $k\cdot p$ methods to fit the band structure from the first principles calculation and found that the edge states along the $(100)$ and $(010)$ directions have different Dirac energy due to C$_4$ symmetry breaking. This is essential to obtain Majorana corner states in D class without concerning the details of the superconducting pairing symmetries and Zeeman form, which can potentially bring advantages in the experimental implementation.

\end{abstract}

\pacs{74.45.+c, 75.70.Tj, 85.25.Cp}
\maketitle

\noindent{\textit{Introduction: }}The hybrid of superconductivity and topological band structure can provide an experimentally accessible platform to achieve the Majorana zero modes (MZMs) \cite{Fu2008}. In the early studies, this hybrid is realized in the superconductor/topological insulator heterostructure through the superconducting proximity effect. The proximity induced superconducting gap is sensitive to the interface of the heterostructure and normally one order of magnitude smaller than the gap in the mother superconductor, which bring various difficulties in experimental verification of MZMs. Thus, it is essential to realize the MZMs in a large gap superconductor without complex heterostructure. On the other hand, iron-based superconductor was originally discovered as the first fully gapped high-Tc superconductors, which has multi-bands at the Fermi level \cite{Hanaguri2010,Li2015,Ge2014,Wang_2012,Song2011,Hsu14262,Wang2015}. Recent studies show that the existence of multi-bands at the Fermi level is helpful for the coexistence of high T$_{\rm c}$ superconductivity and topological band structure in one material without the complex heterostructure\cite{Zhang2019,Zhang2018,Wu2016,Peng2019}. For instance, the zero bias conductance peak is observed at the surface vertex core \cite{Kong2019,Zhu2019,Zhang2018,Liu2018}, following the theoretical prediction \cite{Xu2016,Hosur2011}. Meanwhile, the theoretical studies of high order topological superconductors \cite{Zhang2013,Benalcazar2017a,Langbehn2017,Song2017,Ezawa2018,Shapourian2018,
Zhu2018,Geier2018,Khalaf2018,Schindler2018,Ezawa2018a,Schindler2018a,Ezawa2018b,
Ezawa2018c,Wang2018,Dwivedi2018,Yan2018,Wang2018a,Hsu2018,Pan2018,Wu2019,Zhang2019a} provide the new insight to realize the MZM directly from two dimensional and three dimensional systems without breaking the uniformity of the bulk Hamiltonian. Thus, the implementation of the high order topological superconductors in iron-based superconductors can provide a promising approach to achieve the MZMs in one large gap superconductor and avoiding complex heterostructure.

\begin{figure}[t]
\centerline{\includegraphics[width=1\columnwidth]{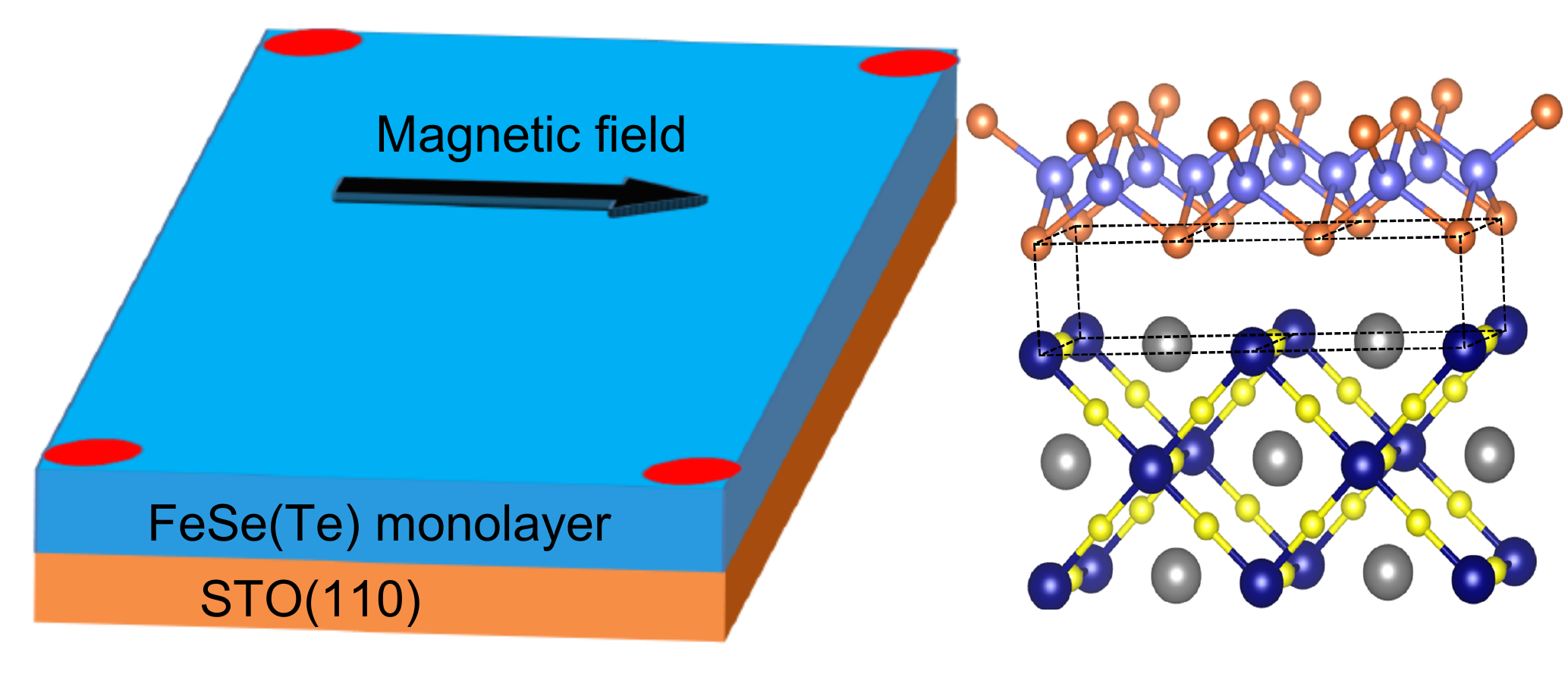}}
\caption{Experimetal set up of FeSe$_x$Te$_{1-x}$/SrTiO$_3$ (110) monolayer. The red points indicate the existence of Majorana corner states. }
 \label{set_up}
\end{figure}

In this work, we theoretically propose to realize the second order topological superconductor and MZMs in the monolayer rectangular FeSe$_x$Te$_{1-x}$/SrTiO$_3$(110). First principles calculations show that the electronic band structure of the monolayer FeSe$_x$Te$_{1-x}$ can be driven from the trivial phase to strong TI phase when one of the in-plane lattice constant is reduced and system symmetry is broken down to D$_{2h}$ in a wide range of composition $x$. This leads to the first order topological phase transition (TPT) of the inverted band structures in the AII class with one pair of helical edge states at each edge. To further considering the topological property in the presence of the superconductivity, we construct the electronic tight-binding model Hamiltonian based on the $k\cdot p$ method with realistic parameters through fitting the bands calculated from the first principles calculations. We note that the C$_4$ symmetry breaking is also necessary for the implementation of Majorana corner states in both time-reversal invariant and breaking monolayer FeSe$_{x}$Te$_{1-x}$. In particular, the edge states along (100) and (010) direction have different electronic Dirac energies, which naturally lead the two edges to be in different gapped phases in the presence of superconductivity and in-plane magnetic field. Our results for D class monolayer rectangular FeSe$_{x}$Te$_{1-x}$/SrTiO$_3$ do not depend on the exact superconducting pairing symmetries (s wave pairing, $s_{\pm}$ pairing ) and the details of the Zeeman form, and are robust against disorders, which lead the rectangle monolayer FeSe$_{x}$Te$_{1-x}$/SrTiO$_3$ to be a promise candidate to realize MZMs without complex hybrid structures.

 \begin{figure}[t]
\centerline{\includegraphics[width=1\columnwidth]{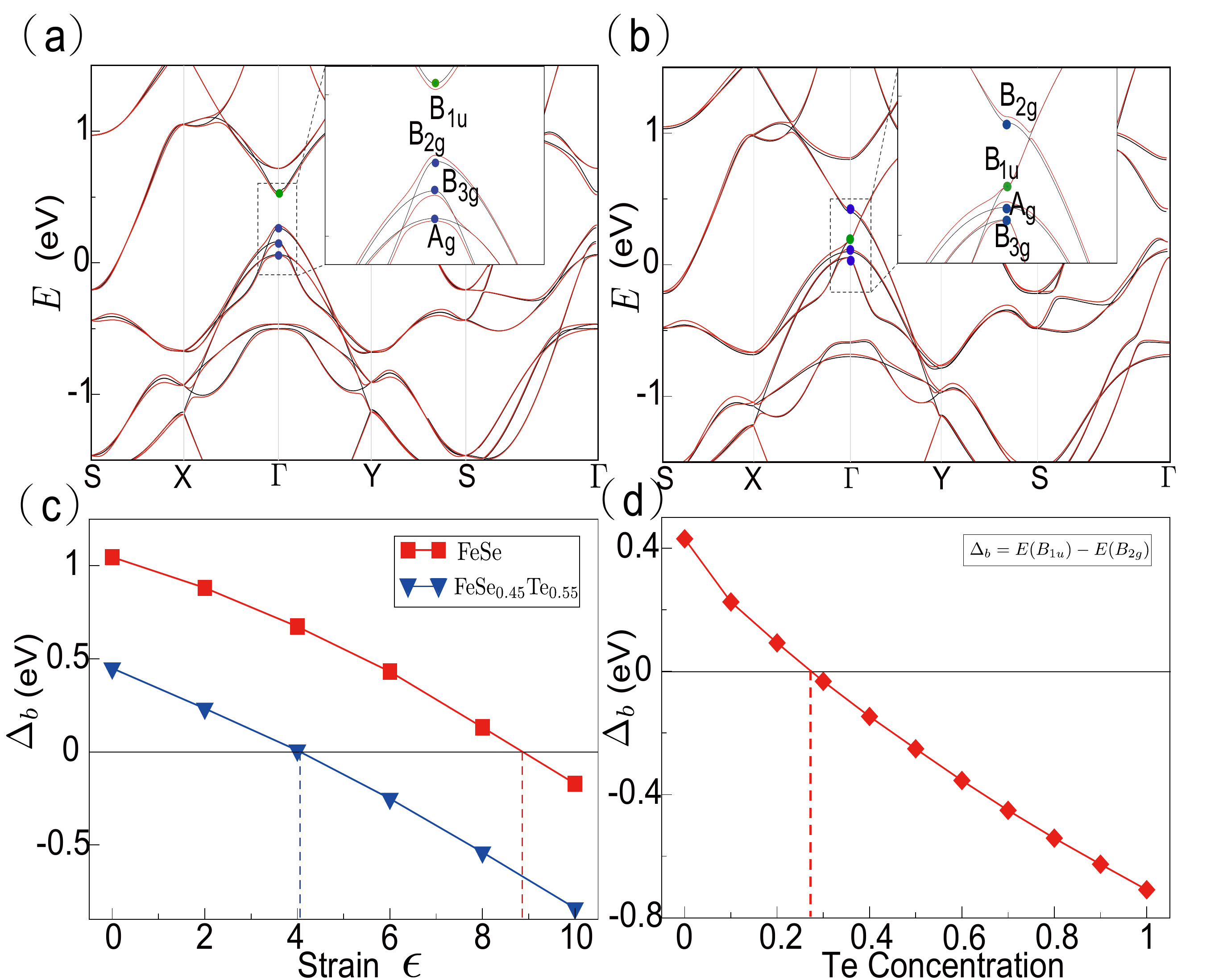}}
\caption{(a) Band structure of FeSe$_{0.45}$Te$_{0.55}$ with the lattice distortion $\epsilon=2\%$.  (b)Band structure of FeSe$_{0.45}$Te$_{0.55}$ with the lattice distortion $\epsilon=6\%$.   (c) The band gap $\Delta_{b}$ of  FeSe$_{x}$Te$_{1-x}$ monolayer as the function of distortion $r$ with $x=1$ (red lines) and $x=0.45$ (blue lines).  (d) The band gap $\Delta_{b}$ of  FeSe$_{x}$Te$_{1-x}$ monolayer as a function of Te concentration $1-x$ with $\epsilon=6\%$. }
 \label{band}
\end{figure}

\textit{Lattice distortion induced First order TPT}: The first principles calculations based on density functional theory (DFT) are carried out to study the topological property of the monolayer FeSe$_x$Te$_{1-x}$ with C$_4$ symmetry breaking. For the experimental reality, the system initially has square lattice with lattice constant $a_0=3.905\rm{\AA}$, which is the same with the (001) monolayer FeSe$_x$Te$_{1-x}$/SrTiO$_3$ \cite{Peng2019}. Without losing the generality, we consider to compress the lattice constant in [100] direction, quantified by the ration $\epsilon=(a_0-a_x)/a_0$ with $a_x$ the lattice constant along x direction after compression. The states at the $\Gamma$ point are thus classified as the D$_{2h}$ representations. Near Fermi level, we consider four orbital states, the odd parity states $B_{1u}$ contributed by Se(Te) $p_z$ orbital and the even parity states $A_{g}$, $B_{2g}$ and $B_{3g}$ mainly coming from Fe $d_{x^2-y^2}$, $d_{xz}$ and $d_{yz}$ orbitals respectively. We distinguish the parities by green and blue dots in Fig.~\ref{band}(a). Note that the $d_{xz}$ and $d_{yz}$ orbitals are not degenerate any more due to the C$_4$ symmetry breaking. We first choose the typical composition value $x=0.45$ \cite{Zhang2018} for which the monolayer FeSe$_x$Te$_{1-x}$/SrTiO$_3$ with C$_4$ symmetry is in topological trivial phase \cite{Peng2019}. The calculated band structures for $\epsilon = 2\%$ and $\epsilon = 6\%$ are plotted in Fig.~\ref{band}(a) and (b) respectively, in which red and black curves correspond to the results calculated with spin-orbital coupling (SOC) and without SOC. For the bands of $\epsilon=2\%$ shown in Fig.~\ref{band}(a), there is no band inversion so that the system is still in topological trivial phase and has a positive band gap at $\Gamma$ point 
$\Delta_{\rm b}=E(B_{1u})-E(B_{2g})>0$. For the bands of $\epsilon=6\%$ shown in Fig.~\ref{band}(b), band inversion happens between the odd parity state $B_{1u}$ and even parity state $B_{2g}$ at $\Gamma$ point, and has a negative  $\Delta_{\rm b}<0$. 
When SOC is excluded (see the black bands in Fig. \ref{band}b), there is a linear band crossing along $\Gamma-Y$ which is absent along $\Gamma-X$ due to the C$_4$ symmetry breaking. When SOC is included, a gap about 25 meV opens at the band crossing so that the system falls into a 2DTI phase around $\Gamma$ point. Thus, the anisotropic lattice distortion, by increasing the compress ratio, induces the first order TPT. In Fig. \ref{band}(c), we plot the band gap $\Delta_{\rm b}$ as a function of the compress ration $\epsilon$ increasing from 0 to 10\%, for different compositions $x=0$ (red square for FeSe) and $x=0.45$ (blue triangle for FeSe$_{0.55}$Te$_{0.45}$), respectively. The calculated results show that, both  systems are initially in the trivial phase without compression ($\epsilon=0$) \cite{Peng2019} and the band gap $\Delta_{\rm b}$ undergoes a sign change at the critical value $\epsilon_{\rm c}=9\%$ and $\epsilon_{\rm c}=4\%$ (indicated by red and blue dashed lines in Fig.~\ref{band}(c)), for FeSe and FeSe$_{0.55}$Te$_{0.45}$ respectively. As reported by Ref. \onlinecite{Zhang2016}, high-$T_c$ superconductivity has already been observed in FeSe/SrTiO$_3$(110) with the rectangular lattice \cite{Zhang2016}. The SrTiO$_3$(110) has the lattice constants $a=3.905\rm{\AA}$ and $b=\sqrt{2}a$. This lattice mismatch makes three unit cells of FeSe grow on the top of two STO(110) unit cells, which reduces the FeSe lattice constant in [100] direction to $\frac{2}{3}\sqrt{2}a \approx 0.94a$, corresponding to  $\epsilon=6\%$ in our calculations. This growth technique should also be applied to FeSe$_x$Te$_{1-x}$ monolayer. We thus take $\epsilon=6\%$ for the experimental reality and plot the band gap $\Delta_{\rm b}$ as a function of the Te composition in Fig.~\ref{band}(d). Such results strongly suggest that both the superconductivity and topological band structures could coexist in the rectangular lattice FeSe$_x$Te$_{1-x}$ for $0.3< x < 0.7$ \cite{Shi2017,Zhang2016,Peng2019}.

\begin{figure}[t]
\centerline{\includegraphics[width=1\columnwidth]{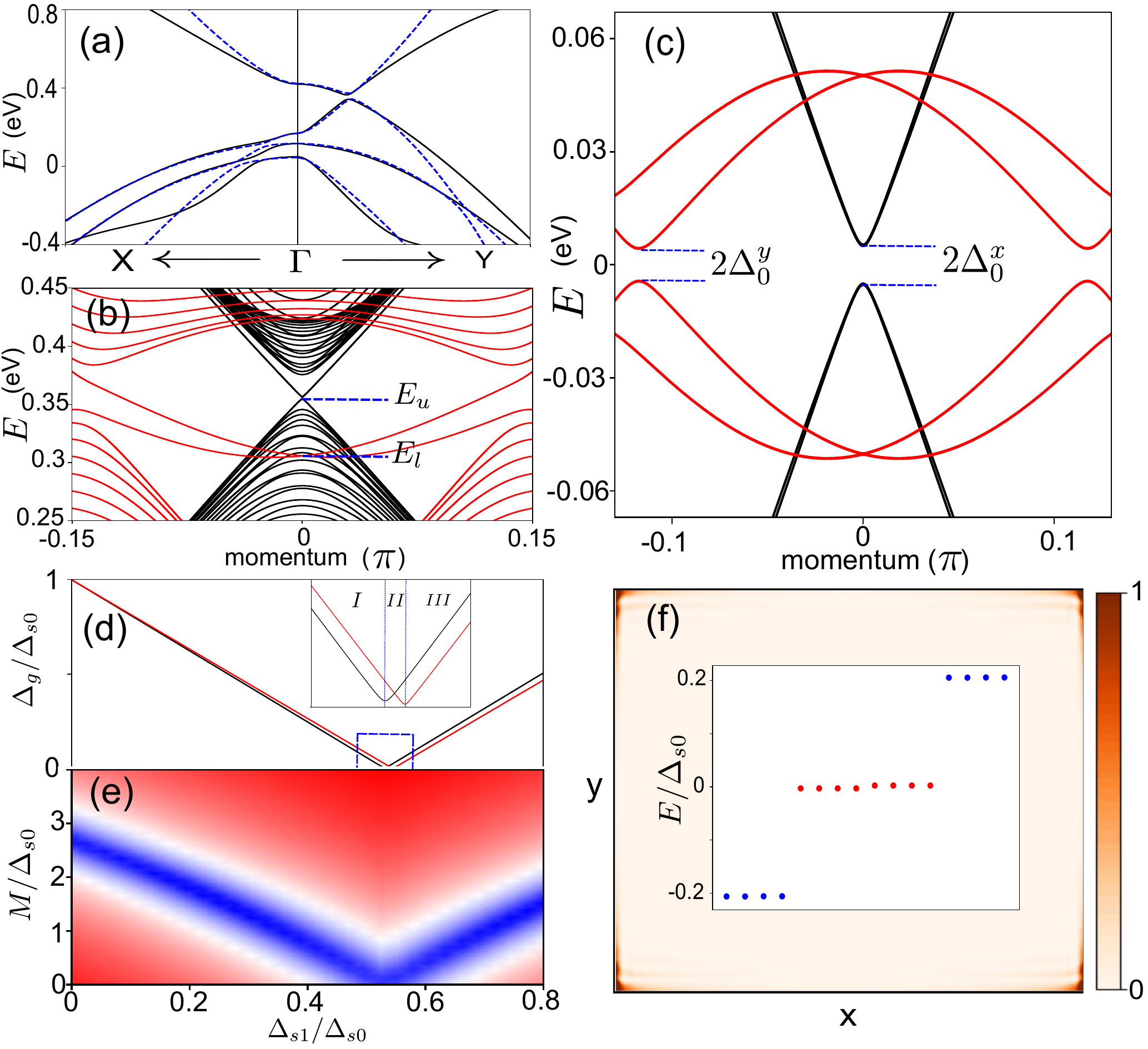}}
\caption{(a) fitting the band dispersions (black) for $\epsilon=6\%$ with $k\cdot p$ method (blue dash). (b) black and red curves for the band dispersions of $k_x$ and $k_y$ based on the fitting parameters in a slab geometry; (c) Band dispersions along x (black) and y (red) in Nambu space;  (d) superconducting gap amplitudes at x (black) and y (red) edge as a function of $\Delta_{s1}$ without Zeeman field. (e) the gap amplitude at x edge as a function of $\Delta_{s1}$ and $M$ (f) The density plot of the MZMs with TR symmetry. The inset plot the several lowest eigenenergies. }
 \label{edge}
\end{figure}

{\it Model Hamiltonian}:
To further investigate the topological edge states, we construct an eight bands $k \cdot p $ effective model Hamiltonian with D$_{2h}$ symmetry  \cite{supp}. The full Hamiltonian with SOC in the basis $ ( |\uparrow \rangle,  |\downarrow \rangle) \otimes ( |yz \rangle , |x^2-y^2 \rangle , |xz\rangle , |z \rangle )$ takes the form $H(\mathbf{k})=H_0\otimes \mathbf{1}_2+H_{soc}$  and the four bands model  $H_{0}$ without SOC has the form:
\beqn\label{Ham}
H_{0}= \left( \begin{array}{cccc}
M_1(\bf{k}) & 0 & \beta k_x k_y & i \gamma k_y \\
0 & M_2(\bf{k}) & 0 & 0 \\
\beta k_x k_y  & 0 & M_3(\bf{k}) & i \delta k_x \\
-i \gamma k_y & 0 & -i\delta k_x & M_4(\bf{k})
\end{array}
\right),
\eeqn
 where $M_{i}(k) = E_i + M_{ix} k_x^2 +M_{iy} k_y^2  $ $(  i=1,2,3,4)$ . The $C_{4}$ symmetry breaking makes $ M_{ix} \neq  M_{iy}$. The parameters of our Hamiltonian are obtained by fitting the bands of the rectangle FeSe$_{0.45}$Te$_{0.55}$ (Fig.~\ref{band}(b)). In Fig.~\ref{edge}(a), we show that, with the fitting parameters, our model can describe the band dispersion near $\Gamma$ point well. The explicit form of the Hamiltonian and fitting parameters can be found in Supplementary Materials \cite{supp}. With these parameters, in Fig.~\ref{edge}(b) we plot the band dispersions for the system in the slab geometry with the open boundary along $x$ (black curves) and $y$ (red curves) respectively. The Dirac points of these two edges have different energies with $E_{u}=0.36$ eV (upper Dirac point) and $E_{l}=0.31$ eV (lower Dirac point). The Dirac energy difference, $\delta E_{\rm D}$ is about 45 meV which is an order of magnitude larger than the superconducting gap and Zeeman splitting energy and consistent with the first principles calculations \cite{supp}. It is convenient to consider the system in Nambu space for including the superconductivity in mean field level. In this case, each edge has two Dirac points near the Fermi level because the degree of freedoms of the system is doubled. Note that as C$_4$ symmetry is broken by lattice distortion, the edge states should be considered separately for $x$ and $y$ edge. For simplicity, we first consider the chemical potential $\mu=E_{u}$, say at the Dirac point of the $x$ edge (black curves in Fig.~\ref{edge}(b)). When the superconducting gap is absent, this Dirac point is doubled at $k_x=0$ and the edge states are four fold degenerate which is protected by both time-reversal and charge U(1) symmetry.  Along $y$ direction (red curves in Fig.~\ref{edge}(b)), the electronic Dirac point at $E_{l}$ is far below the chemical potential so that near the Fermi level there are two separated Dirac points, each of which is two fold degenerate and protected solely by the charge U(1) symmetry \cite{supp}.
 
{\it MZMs in DIII class:} The pairing symmetry of the monolayer FeSe still has many debates in various studies which give plain s-wave pairing \cite{Liu2012b,Fan2015,Fang2011,Zhou2011,Yang2013}, s$_{\pm}$-wave pairing \cite{Stewart2011,Hirschfeld_2011,Mazin2011,Yin2014,Hu2013} and d-wave pairing \cite{Zhang2016a,Agterberg2017,Ge2019,Maier2011,Wang2011}. Recent studies show that when the TI breaks C$_4$ symmetry, both s$_{\pm}$-wave pairing and d-wave pairing can lead to a pair of MZMs at each corner while plain s-wave pairing cannot \cite{Wang2018,Yan2018}. Here, we do not intend to distinguish these pairing symmetries but show that the rectangular monolayer FeSe$_x$Te$_{1-x}$ can potentially be high order time-reversal invariant TSC if it has s$_{\pm}$-wave or d-wave pairings. When applying the s$_{\pm}$-wave pairings $\Delta_{\rm s0}-\Delta_{\rm s1}(\cos(kx)+\cos(ky))$, the charge U(1) symmetries are broken so that the degeneracy at the Dirac points along $x$ and $y$ edges are lifted, which results in two gaps $\Delta_0^{x}$ and $\Delta_0^{y}$ (Fig.~\ref{edge}(c)). The gaps at $x$ and $y$ edges can change sign with varying the gap amplitude $\Delta_{\rm s1}$ for given $\Delta_{\rm s0}$, leading to $\Delta_{0}^x \Delta_0^y<0$ only in regime II (Fig.~\ref{edge}(d)) due to $C_{4}$ symmetry breaking, which is consistent with the results of Ref.~\onlinecite{Yan2018}. We calculate the eigenvalues of the system and found that a pair of MZMs at each corner appear in the regime of $\Delta_{0}^{x}\Delta_{0}^{y}<0$ (Fig.~\ref{edge}(f)). The topological regime is roughly within 0.5 meV around $\Delta_{\rm s0}=2\Delta_{\rm s1}$ for s$_{\pm}$ pairing symmetry according to our fitting parameters. 

\begin{figure}[t]
\centerline{\includegraphics[width=1\columnwidth]{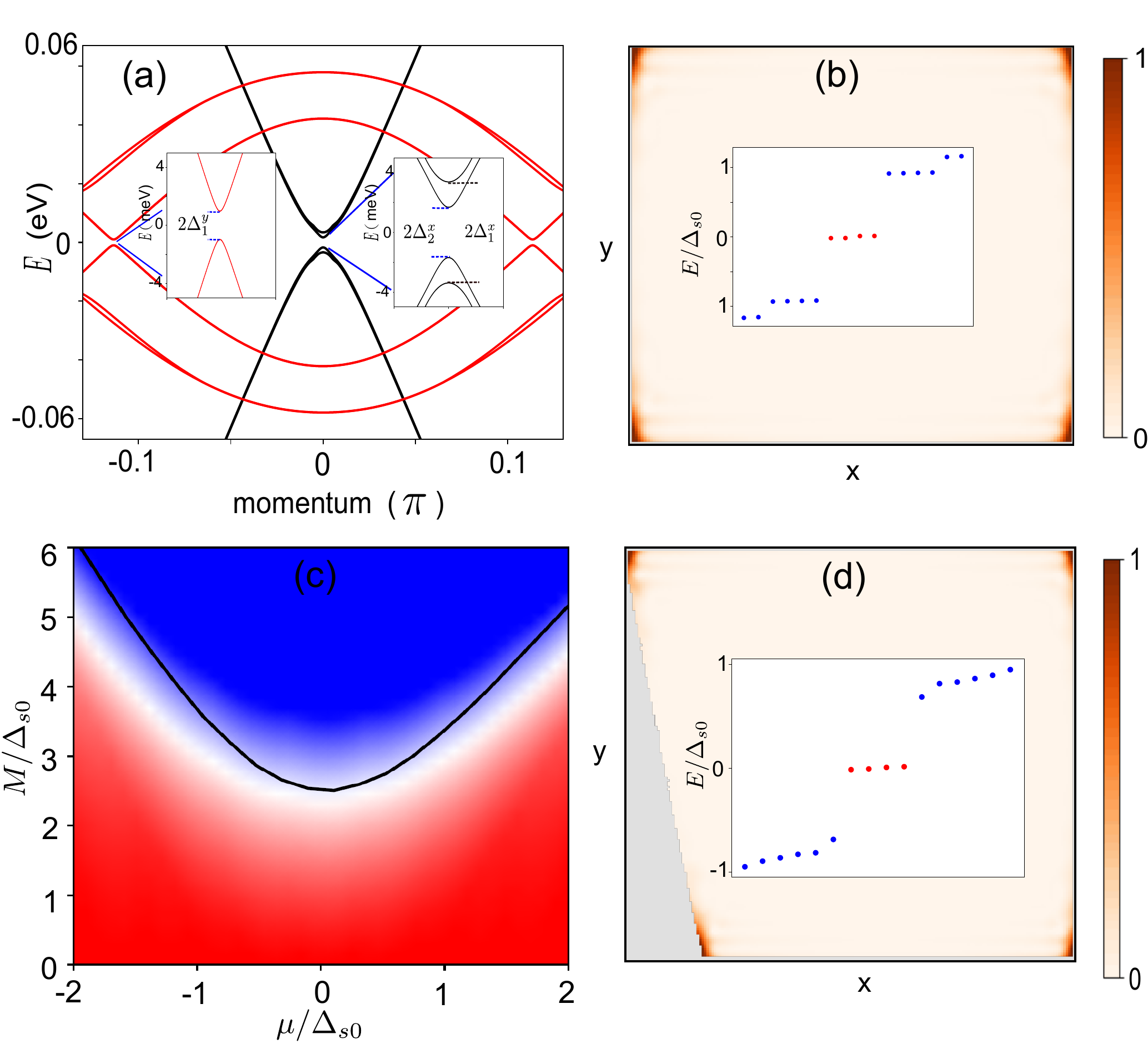}}
\caption{(a)  Band dispersions along x (black) and y (red) in Nambu space with magnetic field.  (b)  The density plot of the MZMs with magnetic field which break TR symmetry. The inset plot the several lowest eigenenergies.  (c) Phase diagram with chemical poential and magnetic field.  (d)  The density distribution and energys of MZMs when the shape of corner is not perfect.  }
 \label{phase}
\end{figure}

{\it MZMs in D class:} When applying a magnetic field induced Zeeman term $M s_x$,  the time-reversal symmetry is broken and the gaps for $x$ and $y$ edges behave in very different manners. Along $x$ direction (Fig.~\ref{phase}(a)), both time-reversal and charge U(1) symmetries breaking results in two gap amplitudes $\Delta_{1(2)}^{x}=\Delta_{0}^x\pm \tilde{M}$ with $\tilde{M}$ the Zeeman splitting of edge states \cite{supp}. As the superconducting gap and the Zeeman term commute, the states at the Dirac points are also the eigenstates of the superconducting matrix $\tau_x s_0$. The state with eigenenergy $-|\Delta_{0}^x-\tilde{M}|$ has the eigenvalue $\nu_{2}^{x}=-1(1)$ for $\Delta_{0}^x-\tilde{M}>(<)0$ respectively while the state with eigenenergy $-|\Delta_{0}^x+\tilde{M}|$ always has the eigenvalue $\nu_{1}^{x}=-1$. Along $y$ direction (Fig.~\ref{phase}(a)), the gap amplitudes are almost independent of the Zeeman term and have $\Delta_{1}^y=\Delta_{2}^y \approx \Delta_{0}^y$ for $\tilde{M} \ll \delta E_{D}$ \cite{supp}. We thus can define a $Z_2$ topological invariant $(-1)^{\nu} = \sign(\nu^{x}_{1}\nu_{2}^{x})$. Based on our model Hamiltonian, the eigenvalues of the system are calculated for both $\nu =0,1$. We found that there are four states with zero eigenvalues, which are MZMs and localized at the four corners for $\nu=1$ (Fig.~\ref{phase}(b)) while the MZMs are absent for $\nu=0$. In Fig.~\ref{edge}(e), we show that to achieve $\nu=1$ has no limit to the ratio $\Delta_{\rm s1}/\Delta_{\rm s0}$. This means the implementation of Majorana corner state in D class monolayer FeSe$_x$Te$_{1-x}$ is not sensitive to the superconducting paring symmetries. Without loss of generality, we take $\Delta_{\rm s1}=0$ in the rest of this work. So far the chemical potential is taken $\mu=E_{u}$. When we vary the chemical potential, the Majorana corner states still exist in a wide chemical potential range. We calculate the lowest eigenenergy of the closed system as function of chemical potential and magnetic field. The color plot of the eigenenergy (Fig.~\ref{phase}(c)) shows an obvious phase boundary between zero (blue) and finite (red) values. We also calculate the critical magnetic field (black curve in Fig.~\ref{phase}(c)), where the gap of the edge state along $x$ direction is closed, as a function of chemical potential. The phase boundary matches the critical magnetic field well, which means the Majorana corner states in our work is not sensitive to the chemical potential as along as it does not close the edge states gap.

As the anisotropic edge states, resulting in the different electronic Dirac point energies, is essential for realize the Majorana corner state, we construct the edge theory to study the effect of the lattice distortion on the edge states.
Because the band inversion at $\Gamma$ point only take place between $|xz\rangle$ and $|z\rangle$ orbitals with the highest two energies, without loss of generality, we can construct our edge theory in a simplified model \cite{Yan2018}
 \begin{align}
 H_{eff}(k)= B(k) s_0 \sigma_z-D(k) s_0 \sigma_0 -\eta k_x s_0 \sigma_y +\alpha k_y s_z \sigma_x,\nonumber
 \end{align}
by projecting the Hamiltonian (Eq.~\eqref{Ham}) into these two orbitals with $D(k)=D_x k_x^2 +D_y k_y^2$ and $B(k)=E_b-B_x k_x^2 -B_y k_y^2$. Note that the term with $D(k)$ breaks the conduction-valence symmetry of the bands while $D_{x} \neq D_{y}$ and $B_{x} \neq B_{y}$ due to the C$_4$ symmetry breaking.

At first we consider the edge states in (100) direction for  the semi-infinite system with $y\in (0, \infty)$. In this case, we
decompose the Hamiltonian as  $H_{eff}=H_{0}+H_{p}$ with
\begin{eqnarray}
H_{0}(k_{x},-i\partial_{y})&=&(E_b+B_y \partial_{y}^{2})s_0 \sigma_z -i \alpha \partial_{y}s_z \sigma_x,\nonumber\\
H_{p}(k_{x},-i\partial_{y})&=&D_y \partial_{y}^{2}s_0 \sigma_0 -\eta k_x s_0 \sigma_y.
\end{eqnarray}
By projecting the Hamiltonian $H_p$ into the basis $\psi_{\alpha=1,2}(y)$, which are the eigenstates of $H_0$, we obtain the effective edge states Hamiltonian
\begin{eqnarray}
H_{x}(k_{x})= \frac{D_y E_{b}}{B_y} \tilde{s}_0 + \eta k_x \tilde{s}_z 
\end{eqnarray}
with $\tilde{s}$ the Pauli matrix acting in edge states space. Similar we can also obtain the edge states Hamiltonian in (010) direction which has the form
\begin{eqnarray}
H_{y}(k_{y})= \frac{D_x E_{b}}{B_x}  \tilde{s}_0 -\alpha k_y \tilde{s}_z.
\end{eqnarray}
We found that the difference of the Dirac energys $E_{u}-E_{l}=E_b(\frac{D_y}{B_y}-\frac{D_x}{B_x})$ which is only finite with breaking both conduction-valence band symmetry ($D_{i} \neq 0$) and C$_4$ symmetry ($B_{x} \neq B_{y}$) spontaneously. The former is naturally satisfied for $|z\rangle$ and $|xz\rangle$ orbitals. With the fitting parameters, we find that $E_{u}-E_{l}=48$ meV calculated from the four band model is consistent with DFT calculations\cite{supp}.

{\it Disucssion and conclusion:}  Considering the experimental reality, the d-orbitals of the iron-based superconductors may have very complicated g factor, which can result in orbital dependent Zeeman term other than $M\sigma_0 s_x$. However, the various Zeeman terms can always lead to the transition from $\nu=0$ to $\nu=1$ \cite{supp}. Thus our results are independent of the special Zeeman forms. The edges of the iron-based superconductor may not be perfectly along $x$ or $y$ directions and the corner maybe not sharp, which however do not affect the robustness of the Majorana corner state due to its protection only from particle-hole symmetry. In Fig.~\ref{phase}(d), the four eigenfunctions with the lowest eigenvlaues are plotted in the system whose two neighbor edges have a angle $105\degree$ with smooth corners. These states localize at the four corners with perfect zero energy which shows the robustness of the MZMs under the imperfect edges and corners.  In conclusion, the C$_4$ symmetry breaking by the lattice distortion in monolayer FeSe$_x$Te$_{1-x}$ can lead to the first order topological phase transition in the wide composition range $x\in(0,0.7)$, which includes the regime for the monolayer FeSe$_x$Te$_{1-x}$ with the critical temperature up to 65K \cite{Shi2017}. This rectangular monolayer FeSe$_x$Te$_{1-x}$ is also a promise candidate to realize Majorana corner states. 

\section*{Acknowledgement}

We would like to thank Chao-Xing Liu, Chen Fang, Ling-Yuan Kong and Yi Zhou for fruitful discussions. G. Xu acknowledges the support of the Ministry of Science and Technology of China (Grant No. 2018YFA0307000), and the NFSC (Grant No. 11874022). X. Liu acknowledges the support of National Key R\&D Program of China (Grant No. 2016YFA0401003) and NSFC (Grant No. 11674114).

\end{document}